\documentclass[twocolumn]{aastex701} 

\usepackage{multirow}

\usepackage{graphics,epsf}
\usepackage[utf8]{inputenc}
\usepackage{amsmath}                % American Mathematical Society package
\usepackage{amsfonts}               % American Mathematical Society fonts
\usepackage{amssymb}                % American Mathematical Society symbol
\usepackage{epsfig}                 % EPS figures
\usepackage{graphicx}               % Required for inserting images
\usepackage{float}
\usepackage{color}
\usepackage{multirow}               % double row table entries

\hypersetup{
    colorlinks=true,
    linkcolor=red,   
    urlcolor=cyan}

\usepackage[colorinlistoftodos]{todonotes}

\newcommand{\km}{{~\rm km}}
\newcommand{\s}{{~\rm s}}

\newcommand{\g}{{~\rm g}}
\newcommand{\G}{{~\rm G}}
\newcommand{\K}{{~\rm K}}

\newcommand{\yr}{{~\rm yr}}

\begin{document}

\title{Stellar black hole binaries from two common envelope evolution phases in triple stellar systems}
%\date{September 2024}

\author[0009-0004-5313-4341]{Lotem Unger}
\affiliation{Department of Physics, Technion - Israel Institute of Technology, Haifa, 3200003, Israel; lotem.unger@campus.technion.ac.il; soker@technion.ac.il}
\email{lotem.unger@campus.technion.ac.il}

\author[0000-0003-0375-8987]{Noam Soker}
\affiliation{Department of Physics, Technion - Israel Institute of Technology, Haifa, 3200003, Israel; 
lotem.unger@campus.technion.ac.il; soker@technion.ac.il}
\email{soker@technion.ac.il}

\begin{abstract}
We propose a triple-star evolutionary channel involving two common envelope evolution (CEE) phases to form close binary black hole (BBH) systems with an average positive effective inspiral spin $\chi_{\rm eff}$ and a tail of systems having $\chi_{\rm eff}<0$, as observed by gravitational wave detectors. $\chi_{\rm eff}$ is the mass-weighted spin of the two merging BHs, and a positive (negative) value indicates an effective spin along (opposite) the orbital angular momentum. The first BH progenitor engulfs a low-mass star during the post-main-sequence evolution. The tertiary star spirals in and spins up the core, which forms the first BH at the first core-collapse supernova (CCSN) explosion. Its spin is along the orbital angular momentum of the inner binary, which can be highly inclined to the outer binary angular momentum. The secondary star later engulfs the BH in a second CEE phase and explodes as a CCSN to form the second BH with a spin that is more aligned with the orbital angular momentum of the two BHs. We use empirically calibrated initial distributions of triple-star systems consisting of two massive stars and impose a hierarchical stability criterion. We compare the predicted ratio of merging BBHs to CCSN explosion rates and find it is up to a factor of 2 larger than the observed rate.  This channel can significantly contribute to the population of observed merging BBHs and can explain their qualitative spin distribution. 
\end{abstract}
   
% ==================================
\section{Introduction} 
\label{sec:intro}
% ==================================

Gravitational-wave LIGO-Virgo-KAGRA (LVK) science runs O1–O4 detected about 300 black hole (BH)-BH merger events (e.g., \citealt{LVK2025GWTC4pop}). These detections promoted a huge theoretical study of the formation channels of binary BH (BBH) progenitors.
Most channels fall into two broad categories (e.g., \citealt{MandelFarmer2022} for a review). (1) Isolated binary systems where the merging BBH interact without the influence of a third body, either via a common envelope evolution (CEE; e.g., \citealt{Kruckowetal2016, Klenckietal2021, Belczynskietal2022, Broekgaardenetal2022, Shahetal2026}), or without a CEE (e.g., \citealt{MandelDeMink2016, Marchantetal2016, VignaGomezetal2021, ShaoLi2022, WangZYQinetal2026}); see \cite{Kapiletal2026} for a recent study. (2) The dynamical interaction channels, where a pure dynamical interaction with a third star or stars in a binary BH bound system  (e.g., \citealt{Zevinetal2019, Zevinetal2021, Kummeretal2025, VignaGomezetal2025, Dorozsmaietal2026, Grishinetal2026}), or with stars and ambient gas (like the disk around a supermassive black hole at the center of galaxies) of two initially unbound BHs 
(e.g., \citealt{Mapelietal2022, Grishinetal2024, Gilbaumetal2025, Moncrieffetal2026,Vaccaroetal2026}). 
Note that although CEE is emphasized here, the prevalence of CEE among BBH progenitors, and indeed whether a CEE phase is required at all to form merging BBHs, remains an open question, as merger-rate constraints alone do not fix the formation pathway (\citealt{Broekgaardenetal2026}).  
Yet, there is a third channel: a triple-star system in which the main interaction is via mass transfer and CEE in a triple (or higher multiple) system, rather than a pure dynamical interaction (e.g., \citealt{Soker2022misalignment}). 

\begin{figure}[]
    \centering
    \includegraphics[width=\columnwidth]{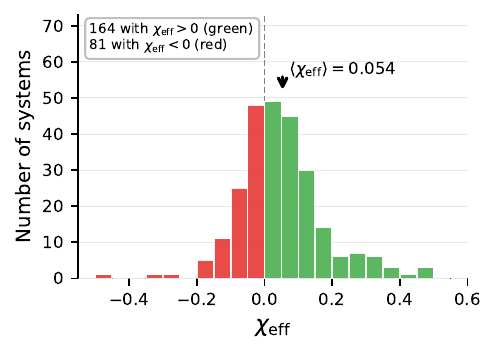}
    \caption{Distribution of the effective spin parameter, $\chi_{\rm eff}$, for 256 BBH events from the O1–O4b observing runs. Asymmetric alignment is observed, with 164 systems possessing positive effective spins ($\chi_{\rm eff} > 0$, green), 81 systems displaying negative spins ($\chi_{\rm eff} < 0$, red) and the 11 systems have zero effective spins ($\chi_{\rm eff} = 0$). The dashed vertical line marks the zero-point, and the black arrow denotes the sample mean ($\langle \chi_{\rm eff} \rangle = 0.054$).}
    \label{fig:chieff_distribution}
\end{figure}

In this study, we address the effective spin of the merging BBHs 
\begin{equation}
\chi_{\rm eff} \equiv \frac{M_1 \vec \chi_1 + M_2 \vec \chi_2}{M_1 + M_2}
\cdot \frac{\vec L} {\vert \vec L \vert} ,
\label{eq:EffectiveSpin}
\end{equation}
where $M_1$ and $M_2$ are the masses of the two BHs, $\chi_1=c J_1/G M^2_1$ and $\chi_2=c J_2/GM^2_2$ are the dimensionless spins of the two BHs, $J_1$ and $J_2$ are the spins of the two BHs, and $\vec L$ is the angular momentum of the orbital motion of the BBH. 
The observed distribution of $\chi_{\rm eff}$ is asymmetric, with more systems having $\chi_{\rm eff}>0$ than those having $\chi_{\rm eff}<0$ (e.g., \citealt{Abbott2023PhRvX, LVK2025GWTC4pop,2026arXiv260527226T}; see Figure \ref{fig:chieff_distribution}). 
% FFFFFFFFFFFFFFFFFFFFFFFFFFFFFFFFFFFFFFF
% FFFFFFFFFFFFFFFFFFFFFFFFFFFFFFFFFFFFFFF

The asymmetric distribution shows that the orbital motion influences the final BH spins and disfavors a purely dynamical origin (e.g., \citealt{Rouletetal2021}). The significant number of systems with $\chi_{\rm eff}<0$ shows that other effects determine the BH spins in addition to the orbital angular momentum. The dynamical channels, where the spins of the two BHs need not be correlated with the final angular momentum of the two merging BHs, predict a large number of systems with $\chi_{\rm eff} \leq 0$. However, they do not account for the large asymmetric distribution of $\chi_{\rm eff}$. 
The expectation that dynamical assembly isotropizes the spins has, however, recently been questioned: \cite{Martinezetal2026} find that BBHs entering with aligned spins can retain that alignment over several encounters, though this inherited alignment adds to the positive $\chi_{\mathrm{eff}}$ side rather than the negative tail. 

One of several processes might impart negative effective spins to BHs in isolated binary system channels without involving purely dynamical effects. 
\begin{enumerate} 
\item A very large natal velocity kick to the second-born BH (e.g., \citealt{Fragioneetal2021}) such that the orbit is flipped. Other observations suggest that BHs do not often acquire the required large kick velocity of $> 150 \km \s^{-1}$, although some might (e.g., \citealt{NagarajanElBadry2025}). 
\item The tossing of the BH spin during its formation (e.g., \citealt{Tauris2022, Larsenetal2025NewA}). \cite{Larsenetal2025NewA} find that BH natal kick alone cannot explain the $\chi_{\rm eff}$ distribution. Either spin-axis tossing is needed, or $72 \pm 8\%$ of merging BBHs result from dynamical interaction channels. 
\item When a triple stellar system experiences two CEE phases, or one CEE and one mass transfer phase (\citealt{Soker2022misalignment}). \cite{Soker2022misalignment} considered the entrance of the tight binary system, including one or two BHs, into the giant envelope of a red supergiant (RSG) of a third star. Although this is a triple-system interaction, it is not purely dynamical (hence it does not belong to the dynamical interaction channels and does not involve the von-Zeipel-Lidov-Kozai instability). The channel we study here differs from those studied by \cite{Soker2022misalignment}, but it does involve double-CEE evolution.  
\end{enumerate}

The $\chi_{\rm eff}$ distribution and spin-orbit misalignment of BBHs are open questions that force different scenarios to make specific assumptions, either about the physical processes or about the fraction of BBHs arising from different channels.  
\cite{Rodriguezetal2016ApJ2} conclude that only dynamical processes can form significant spin-orbit misalignment in BBH mergers. \cite{Baveraetal2020} studied the spin of BBH mergers in the CEE channel of binary systems, assuming that the spin of the first-born BH is very low. They find that most of their BBHs merge with a spin parameter of $\chi_{\rm eff} \simeq 0$, and a few with positive $\chi_{\rm eff}$.  \cite{Belczynskietal2020} consider that BHs are born with very low spin in the framework of the isolated binary channel. 
\cite{Klenckietal2021} find that the CEE requires that the first-born BH enter a CEE with an RSG; the large envelope of RSG stars ensures that the envelope binding energy is sufficiently low to allow successful envelope ejection. \cite{Zevin2021B} argued that a mixture of channels is strongly preferred over any single channel of BBH formation. \cite{Mapelietal2022} find from their analysis that multiple channels contribute to the BBH population. They did not study the origin of spins in isolated BBH, but rather introduced a random distribution as an input parameter. 
Recent data-driven analyses of the GWTC-4 catalog likewise favor more than one subpopulation: \cite{PadhyegurjarMukherjee2026} find the strongest evidence for two channels, with only mild support for an additional AGN-disc component, while \cite{Padhyegurjaretal2026} identify three subpopulations in the delay-time distribution whose local merger rates differ by about an order of magnitude.

In this study, we consider an isolated triple-star channel within the jittering-jets explosion mechanism (JJEM) of core-collapse supernovae (CCSNe). 
\cite{Willcoxetal2026} study the formation of BBH taking their masses in the framework of the neutrino-driven mechanism. We, on the other hand, build our scenario within the JJEM framework. 
\cite{MorenoMendezetal2023} study a triple star evolution with one CEE phase to explain BBHs in the pair-instability-supernova mass gap. We will consider two CEE phases to also explain lower-mass BBHs.   
We describe this scenario for BBH formation in the framework of the JJEM and our assumptions in Section \ref{sec:assumptions}. We then estimate the formation rate of such systems relative to that of all CCSNe; we describe the calculation method and the relevant assumptions in Section \ref{sec:Numerics}, and the results in Section \ref{sec:Fraction}.  We summarize our results and discuss the formation of misaligned BBHs in the context of the scenario we study in Section \ref{sec:Summary}.

% =========================
\section{Basic assumptions of the scenario}
\label{sec:assumptions}
% =========================

We conduct this study in the framework of the JJEM, where a CCSN leads to the formation of a BH when the progenitor is a massive star with a rapidly rotating core. The reason is that when the pre-collapse core rotation is fast, the accretion disk around the newly born neutron star, and then the BH, has a roughly constant axis; some small jittering or precession around the pre-collapse rotation axes is possible. Jets with a fixed axis are less efficient at expelling core material from the equatorial plane vicinity (the disk plane) than jittering jets in different directions (e.g., \citealt{Gilkisetal2016}). If, in addition, the binding energy of the core is large, which implies a massive progenitor, the falling material accumulates to form a BH. This event can lead to a very energetic CCSN, rather than to a failed CCSN (e.g., \citealt{Gilkisetal2016, Soker2026Failed}).   

Our basic assumption, which is fully compatible with the JJEM, is that for the formation of a BH, the progenitor should be massive, a zero-age main-sequence (ZAMS) mass of $M_{\rm ZAMS} \gtrsim 20 M_\odot$, and its inner regions should be spun-up during its post-main-sequence evolution.  
Close binary interaction is required to spin-up the pre-collapse core to fast rotation (e.g., \citealt{Muller2023spin}).  

\begin{figure}[]
    \centering
    \includegraphics[trim=0.9cm 1.4cm 0.0cm 1.2cm ,clip, angle=0, scale=0.85]{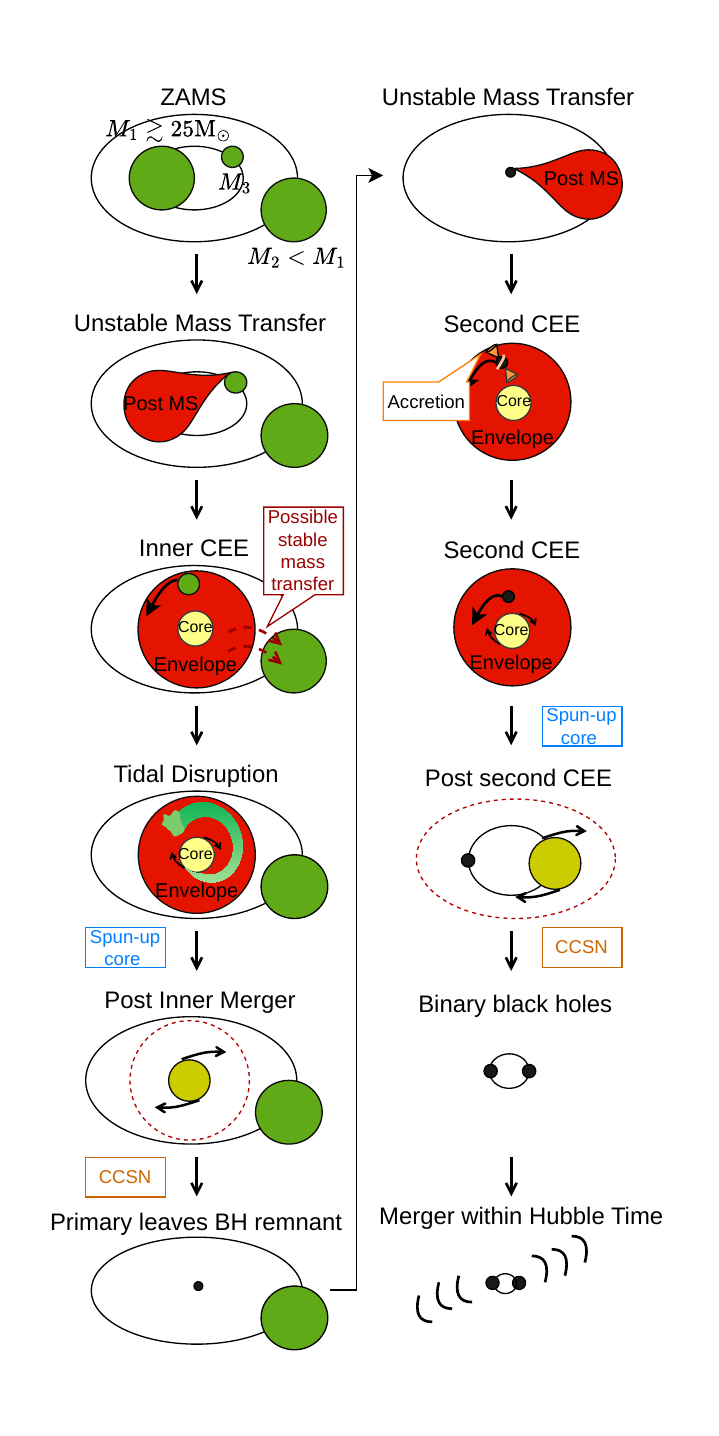}
    \caption{Schematic evolutionary pathway of a hierarchical triple system leading to the formation of a merging BBH. The left column follows the primary star from the ZAMS through the inner CEE, the inner merger (tidal disruption of the tertiary), and the first CCSN, which produces the first-born BH. The right column follows the subsequent evolution of the secondary star through the second CEE and second CCSN, resulting in a BBH that merges within a Hubble time. The individual steps are described in section \ref{sec:assumptions}. The arcs in the final panel schematically represent GW emission; their orientation in the figure is for visual clarity and does not reflect the actual emission geometry.}
    \label{fig:evolution_channel}
\end{figure}

\begin{figure*}[]
    \centering
    \includegraphics[width=0.95\textwidth]{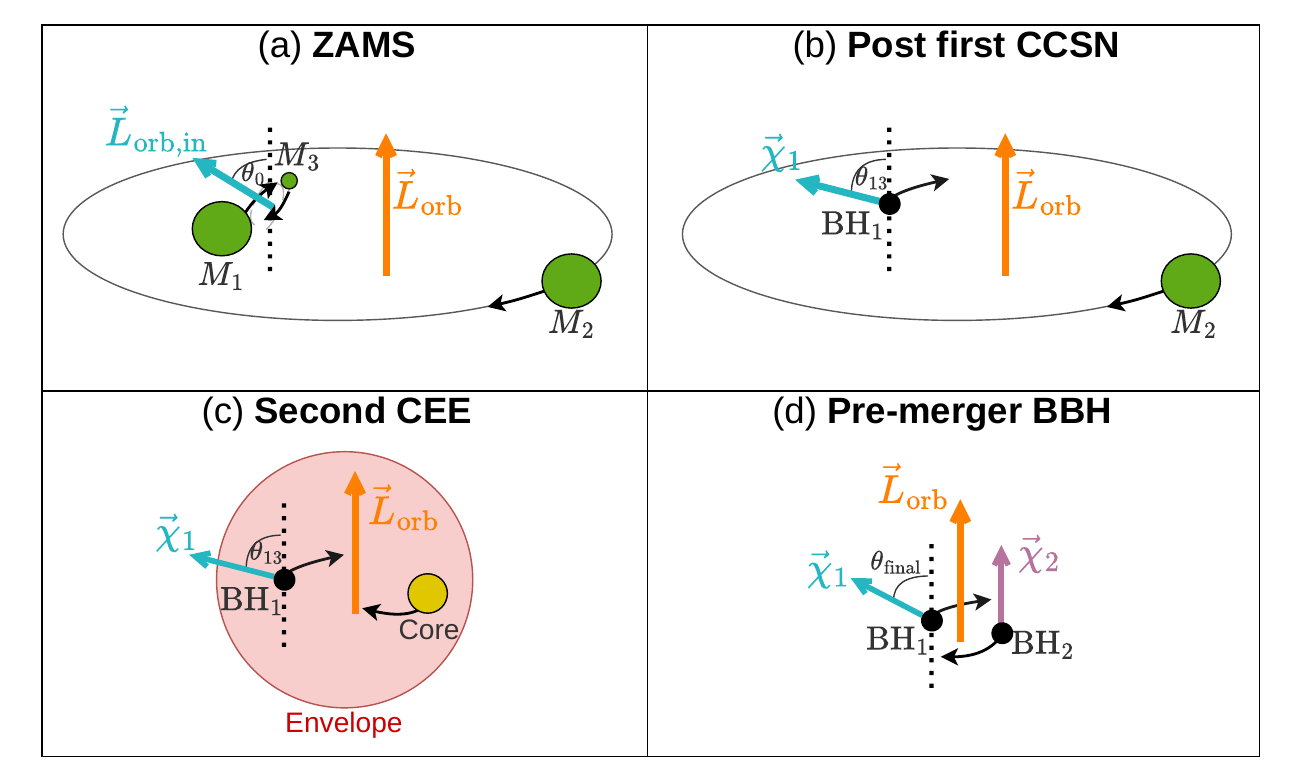}
    \caption{Schematic illustration of the angular momentum and spin configuration at four stages of the evolution described in section \ref{sec:assumptions}. (a) ZAMS: the inner orbital angular momentum $\vec{L}_{\mathrm{orb,in}}$ is inclined by an angle $\theta_0$ relative to the outer orbital angular momentum $\vec{L}_{\mathrm{orb}}$. (b) Post first CCSN: the first-born BH carries a spin $\vec{\chi}_1$ approximately aligned with $\vec{L}_{\mathrm{orb,in}}$, and therefore inclined by $\theta_{13}$ relative to $\vec{L}_{\mathrm{orb}}$; this angle differs in general from $\theta_0$ due to the inner merger and the first CCSN (see text). The thin black arrow emerging from the BH indicates the direction of its orbital velocity. (c) Second CEE: the first-born BH spirals inside the secondary's envelope. (d) Pre-merger BBH. For visual simplicity, $\vec{\chi}_1$ and $\vec{L}_{\mathrm{orb}}$ are drawn with the same orientations as in panel (c); in practice, both may change during the second CCSN, as discussed in the text.}
    \label{fig:am_geometry}
\end{figure*}

We consider a system of two very massive main-sequence stars, and a tertiary star of mass $M_3 \lesssim 5 M_\odot$, orbiting the more massive of the two. 
The evolutionary scenario is depicted schematically in Figure \ref{fig:evolution_channel}, and the corresponding angular momentum geometry is illustrated in Figure \ref{fig:am_geometry}.
In the scenario we study, this more massive star with ZAMS mass of $M_{\rm 1,ZAMS}$, which is the progenitor of the first BH to be born, engulfs the tertiary star as it evolves off the main-sequence. This forms the first CEE phase of the system. The tertiary star spirals all the way to the core vicinity and spins up the core. The core destroys the tertiary star by tidal disruption. The explosion of this star forms a BH with its spin axis $\vec \chi_1$, more or less along the orbital angular momentum $\vec {L}_{\mathrm{orb,in}}$ of the primary and tertiary (inner binary) system. Mass transfer from the RSG primary to the secondary is likely to occur.  
% FFFFFFFFFFFFFFFFFFFFFFFFFFFFFFFFFFFFFFFFFFFFFFFFFFFFFFFFFFFFF
%trim={<left> <lower> <right> <upper>}
% 1. Evolution Channel (Single Column)
% FFFFFFFFFFFFFFFFFFFFFFFFFFFFFFFFFFFFFFFFFFFFFFFFFFFFFFFFFFFFF
% FFFFFFFFFFFFFFFFFFFFFFFFFFFFFFFFFFFFFFFFFFFFFFFFFFFFFFFFFFFFF
%trim={<left> <lower> <right> <upper>}
% 2. Angular Momentum (Two Columns)
% FFFFFFFFFFFFFFFFFFFFFFFFFFFFFFFFFFFFFFFFFFFFFFFFFFFFFFFFFFFFF

When the secondary star, of ZAMS mass $M_{\rm 2,ZAMS} < M_{\rm 1,ZAMS}$ evolves to become an RSG star, it engulfs the first BH, now of mass $M_1$. 
This is the second CEE phase of the system. Note that because of mass transfer from the primary to the secondary, the secondary mass at this stage is likely to be $M_2>M_{\rm 2,ZAMS}$, and might even be $M_2>M_{\rm 1,ZAMS}$.   
The BH spirals close to the core of the secondary star and spins up the zones around the secondary stellar core. The massive, rapidly rotating core forms a BH according to the JJEM; this is the second-born BH. After the explosion of the secondary star, it leaves a remnant, a BH with a spin axis more or less along the orbital angular momentum axis of the binary system. 

The first-born BH is likely to accrete some mass from the envelope. The angular momentum of this accretion process is aligned with the orbital angular momentum of the binary system. Therefore, the accretion process in the second CEE phase decreases the angle between $\vec \chi_1$ and that of the binary orbital angular momentum $\vec{L}_{\mathrm{orb}}$. \cite{Qinetal2018}, who studied binary CEE, argued that the first-born BH has a small spin ($<0.1$). We follow them and assume that the change in the spin of the first-born BH during the CEE is small, although in our scenario its spin does not need to be small.    

Like in earlier studies of similar evolutionary routes (e.g., \citealt{Qinetal2018, Baveraetal2022}),  our scenario predicts that the spin of the second-born BH, $\vec \chi _2$, tends to be aligned with the orbital angular momentum of the BBH, $\vec{L}_{\mathrm{orb}}$, and might be large. However, several processes can cause misalignment: (1) A natal kick velocity of the second-born BH (see Section \ref{sec:intro}) causes the new orbital angular momentum axis, $\hat L$, to differ from the axis of the orbital angular momentum before the explosion, $\hat L \ne \hat L_{\mathrm{orb}}$. (2) Accretion of stochastic angular momentum from the envelope, as the JJEM implies, might cause small changes in the spin axis $\chi_1$. (3) A fourth star in the system that orbits the secondary. The fourth star must be light in order not to prevent the secondary from becoming an RSG. The three processes can cause a misalignment, but not a large one. 

To summarize, the scenario we study in the framework of the JJEM requires a triple-star system to form a BBH. The scenario allows for $\chi_{\rm eff} < 0$, and predicts that it results from the misalignment of the first-born BH, i.e., the axes of $\vec \chi_1$ and $\vec L_{\mathrm{orb}}$ can be highly misaligned, and even opposite. The second-born BH spin tends to be aligned with the orbital angular momentum axis. It also predicts an asymmetrical distribution of $\chi_{\rm eff}$, as it is much more likely to have a positive than a negative value.

% =========================
\section{The numerical settings}
%\section{Calculation method}
\label{sec:Numerics}
% =========================

% =========================
\subsection{Outline of the semi-analytical approach}
\label{subsec:outline}
% =========================

Our goal is to assess whether the triple-star channel of Section \ref{sec:assumptions} can plausibly contribute to the population of merging BBHs, rather than to quantify its precise contribution. Quantifying the contribution would require a population-synthesis treatment of triple systems, which is computationally expensive; we instead adopt a semi-analytical estimate that is sufficient to test whether the channel is a viable contributor.

\begin{figure}[]
    \centering
    \includegraphics[trim=0.2cm 0.6cm 0.0cm 0.0cm ,clip, angle=0, scale=0.41]{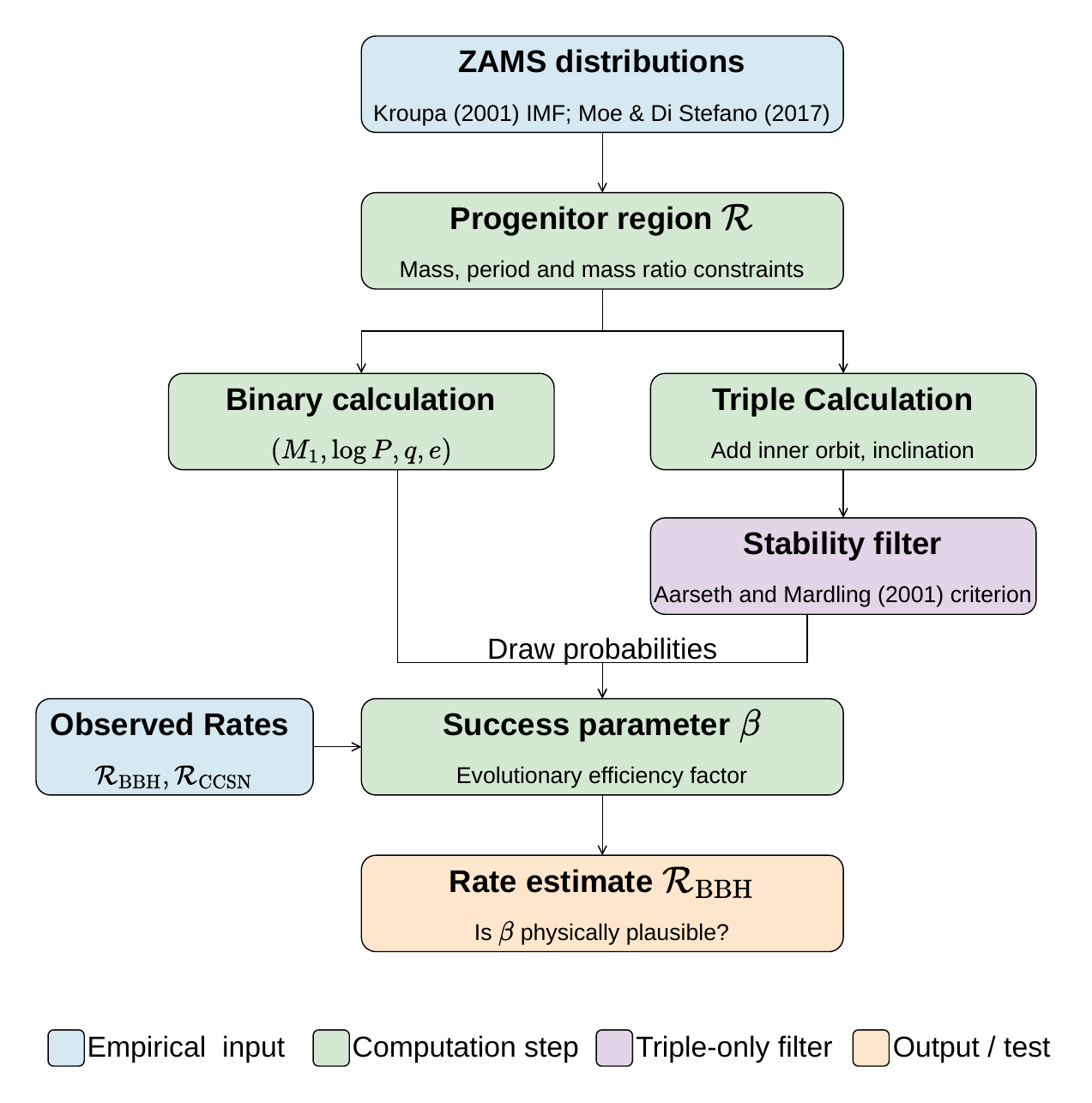}
    \caption{Schematic overview of the analytical calculation pipeline. Blue boxes indicate empirical inputs; green boxes indicate computational steps; the purple box marks the dynamical stability filter applied only to triple systems. The observed CCSN and BBH merger rates are used as external calibrations at the success fraction step. The final output is the inferred success fraction (or evolutionary efficiency factor) $\beta$, whose physical plausibility ($0 < \beta < 1$) serves as the primary test of the formation channel.}
    \label{fig:pipeline_flowchart}
\end{figure}

Because not every triple is born with the properties needed to begin, let alone complete, this evolutionary pathway, the calculation separates into two ingredients. (i) A draw probability $P_{\rm draw}$: the probability that a zero-age main-sequence (ZAMS) system is born with parameters consistent with the configuration the channel requires. (ii) A success fraction $\beta$: the conditional probability that a system satisfying these ZAMS conditions evolves through the full sequence - passes two CEE phases, two CCSNe each forming a BH, and a final orbit tight enough to merge within a Hubble time. The factor $\beta$ absorbs the cumulative uncertainties of the scenario, such as the CEE ejection efficiency, the natal kicks, the mass loss, and the requirement of a sufficiently tight post-collapse orbit. The logical structure of the calculation is summarized schematically in Figure \ref{fig:pipeline_flowchart}.
% FFFFFFFFFFFFFFFFFFFFFFFFFFFFFFFFFFFFFFFFFFFFFFFFFFFFFFFFFFFFF
%trim={<left> <lower> <right> <upper>}
% 3. Flowchart Pipeline (Single Column)
% FFFFFFFFFFFFFFFFFFFFFFFFFFFFFFFFFFFFFFFFFFFFFFFFFFFFFFFFFFFFF

We normalize the calculation to the CCSN rate - an observed quantity. The local BBH merger-rate density is then
\begin{equation}
\mathcal{R}_{\rm BBH} = \frac{\mathcal{R}_{\rm CCSN}}{f}\, P_{\rm draw}\, \beta ,
\label{eq:rate}
\end{equation}
where $\mathcal{R}_{\rm CCSN}$ is the volumetric CCSN rate and $f$ is the mean number of stars massive enough to undergo a CCSN per massive-primary system. The factor $f$ converts the event rate $\mathcal{R}_{\rm CCSN}$ into a rate of progenitor systems: because massive stars reside predominantly in binaries and higher-order multiples (\citealt{Sanaetal2012, MoeDiStefano2017}), a system whose primary explodes as a CCSN typically hosts at least one further companion massive enough ($M \gtrsim 8\,M_\odot$) to do the same. In the channel studied here, both the primary and the secondary collapse to BHs, and averaging the companion statistics over the primary mass function gives $f \simeq 2$. 
We adopt a local CCSNe rate of $\mathcal{R}_{\rm CCSN} \simeq 6.9 \times 10^{4}~{\rm Gpc^{-3}\,yr^{-1}}$ (\citealt{Maetal2025}). \cite{2026arXiv260527226T} gives a BBH merger rate of $27.5-49.4~{\rm Gpc^{-3}\,yr^{-1}}$ at $z=0.2$, we convert this value to $z=0$ to ensure consistency with the local CCSNe rate (see appendix \ref{app:rate_conversion}). This conversion yields a local BBH merger rate of $\mathcal{R}_{\rm BBH} \simeq 17.4-31.3~{\rm Gpc^{-3}\,yr^{-1}}$, which is consistent with the local GWTC-4.0 value of $14-26~{\rm Gpc^{-3}\,yr^{-1}}$ (\citealt{LVK2025GWTC4pop}). Using these values results in a rate ratio of $\mathcal{R}_{\rm BBH}/\mathcal{R}_{\rm CCSN} \simeq 3.4 \times 10^{-4}$, allowing equation \ref{eq:rate} to be rearranged into the form (with $f=2$): 
\begin{equation}
\beta = f\,\frac{\mathcal{R}_{\rm BBH}/\mathcal{R}_{\rm CCSN}}{P_{\rm draw}}
       \simeq \frac{2 \times \left(3.4 \times 10^{-4}\right)}{P_{\rm draw}} .
\label{eq:beta}
\end{equation}

This relation is the central test of the channel: for a computed $P_{\rm draw}$, the inferred $\beta$ must satisfy $0 < \beta < 1$ to be interpretable as a conditional probability. A value exceeding unity would mean that the adopted ZAMS configuration is too restrictive to accommodate the observed BBH rate through this channel alone, whereas a value comfortably below unity indicates that the channel can supply a meaningful share of the rate without requiring an implausibly large fraction of progenitor systems to succeed.

% =========================
\subsection{ZAMS distributions and the draw probability}
\label{subsec:zams}
% =========================

To evaluate the draw probability, we specify the joint ZAMS distribution $f$ and the region of parameter space over which it is integrated,
\begin{equation}
P_{\rm draw} = \int_{\mathcal{R}} f(M_1, \log P, q, e)\; dM_1\, d\log P\, dq\, de ,
\label{eq:Pdraw}
\end{equation}
where $\mathcal{R}$ is the region picked out by the cuts on these parameters. The specific cuts and their physical justifications are presented alongside the results in section \ref{sec:Fraction}. 
For consistency, all subsequent mass and temporal quantities are reported in solar masses ($M_\odot$) and days, respectively.

For the joint distribution, we adopt the empirically calibrated massive-star statistics of \cite{MoeDiStefano2017}, who compile and de-bias multiple observational samples and provide joint fits of primary mass, orbital period, mass ratio, and eccentricity suitable for population-synthesis initialization. Following \cite{MoeDiStefano2017} and \cite{Stegmannetal2022_evolution}, we represent the joint distribution using conditional factors that mirror the structure of the empirical fits,
\begin{equation}
\begin{split}
    f(M_1, & \log P, q, e) \simeq  f(M_1) f(\log P \mid M_1) \\
    & \times f(q \mid \log P, M_1) f(e \mid \log P, M_1) \\
    & \times \Pi_{\mathrm{occ}}(M_1, \log P) ,
\end{split}
\label{eq:prob_expression}
\end{equation}
where $\Pi_{\mathrm{occ}}$ is a companion occurrence factor - the probability that a system with primary mass $M_1$ has at least one companion in the period range being sampled.

The primary mass follows the standard initial mass function (\citealt{Kroupa2001}) described in the high mass regime by the power law $\xi(M) \propto M^{-2.3}$. To remain consistent with the CCSN normalization of equation \ref{eq:rate}, the parent population is defined as all stars capable of core collapse ($M_1 \geq 8\,M_\odot$), so every reported draw probability is conditioned on $M_1 \geq 8\,M_\odot$.

Because we impose no eccentricity cut beyond the requirement that the system not overflow its Roche lobe at periastron, the eccentricity factor $f(e \mid \log P, M_1)$ evaluates to unity over the sampled range. The \cite{MoeDiStefano2017} mass-ratio calibration is a continuous three-regime piecewise fit defined for $q \geq 0.1$; since our scenario requires a low-mass tertiary (Section \ref{sec:assumptions}) whose mass ratio $M_3/M_1$ can fall below this floor, we extrapolate the distribution down to $q_{\mathrm{min}} = 0.05$ for that companion, and find the resulting change in $P_{\rm draw}$ to be small.

We verified that $P_{\rm draw}$ is not strongly sensitive to these choices: replacing the \cite{MoeDiStefano2017} period, mass-ratio, and eccentricity prescriptions with the simplified priors common in rapid population synthesis (log-uniform in period, uniform in mass ratio, and uniform or thermal in eccentricity) yields quantitatively similar values.

% =========================
\subsection{Extension to hierarchical triples}
\label{subsec:triples}
% =========================

For triples, we extend the parameter space to include the inner orbit and the mutual inclination, $(M_1, \log P_{\rm in}, q_{\rm in}, e_{\rm in};\ \log P_{\rm out}, q_{\rm out}, e_{\rm out};\ i)$, where "in" denotes the inner (primary-tertiary) orbit and "out" the outer orbit of the secondary about the inner pair. The mutual inclination angle $i$ is sampled uniformly in $\cos i \in [-1,1]$, the maximum-entropy choice corresponding to isotropic relative orientations.   
%Following the ordering introduced in Section \ref{subsec:outline}, the triple draw probability factorizes as
\begin{equation}
P_{\rm draw}^{\rm triple} \approx P_{\rm out} \times P_{\rm in} \times P_{\rm stable},
\label{eq:triplefactor}
\end{equation}
where $P_{\rm out}$ is the probability that a massive primary has a companion satisfying the outer-orbit cuts - identical to the binary draw probability of equation \ref{eq:Pdraw}. $P_{\rm in}$ is the conditional probability that such a system additionally hosts a tertiary satisfying the inner-orbit cuts. The last factor $P_{\rm stable}$ is the fraction of the resulting configurations that are dynamically stable.

For the stability of the hierarchy, we adopt the criterion of \cite{MardlingAarseth2001} with the inclination-dependent factor of \cite{AarsethMardling2001}, imposed as a hard cut on the ZAMS parameter space:
\begin{eqnarray}
\begin{split}
\left(\frac{a_{\rm out}}{a_{\rm in}}\right)_{\rm crit} & = 
\frac{2.8}{1 - e_{\rm out}}\left(1 - \frac{0.3\,i}{\pi}\right) \\
& \times 
\left[\frac{(1 + q_{\rm out})(1 + e_{\rm out})}{(1 - e_{\rm out})^{1/2}}\right]^{2/5},
\label{eq:stability}
\end{split}
\end{eqnarray}
a configuration being retained only if its semi-major-axis ratio exceeds this critical value. Here $q_{\rm out} \equiv M_2/(M_1 + M_3)$ is the ratio of the outer (secondary) mass to the total mass of the primary-tertiary inner binary.

% =========================
\subsection{Dynamical realizability with population synthesis}
\label{subsec:mse}
% =========================

Equation \ref{eq:stability} is evaluated at ZAMS and is therefore only a snapshot: it does not guarantee that a formally stable triple remains stable as the stars evolve and the orbits change, nor that it actually proceeds through the sequence of section \ref{sec:assumptions}. To verify the dynamical realizability of the channel, and in particular to establish the ZAMS period ranges over which it proceeds, we evolve the sampled triples with the Multiple Stellar Evolution (MSE) population-synthesis code (\citealt{Hamersetal2021}). MSE switches dynamically between an orbit-averaged (secular) treatment and direct $N$-body integration, so that the hierarchical-stability condition is re-evaluated continuously rather than only at ZAMS, and couples this dynamics to single-star evolution from the SSE tracks of \cite{Hurleyetal2000} and to mass-transfer and CEE prescriptions adapted from BSE (\citealt{Hurleyetal2002}).

We use these runs as a realizability check rather than as a rate prediction. The essential reason is that MSE assigns compact remnants from the SSE fitting formulae on the basis of mass alone and does not follow the rotational state of the core; it therefore cannot represent the step that defines the present channel -- the spin-up of the primary's core by the inspiralling tertiary that the JJEM requires for BH formation (section \ref{sec:assumptions}) - and forms BHs whether or not the core was spun up, so that its BH yields are permissive relative to the full scenario. Secondary prescription choices bias the simulated yields in both directions: the mass-only remnant criterion is permissive, whereas the CEE energy budget credits only the orbital-energy term and omits additional sources (for example, energy injected by the accreting BH during the second CEE), which would ease envelope ejection for a wider range of systems. The simulations are accordingly best read as confirming that the dynamical sequence is realizable, and as localizing the inner-orbit period range over which the channel proceeds (section \ref{sec:Fraction}), not as predicting its absolute rate.

% =========================
\section{The fraction of black hole binaries}
\label{sec:Fraction}
% =========================

% =========================
\subsection{Binary calculation}
\label{subsec:BinaryResults}
% =========================

Using the ZAMS prescription described in \ref{subsec:zams}, we first evaluate the analytic draw probability for systems capable of forming merging BBHs, treating the primary and the massive companion as an isolated interacting binary. The integration region is defined by imposing the following constraints on the ZAMS parameters.
 
\textbf{Primary mass.} We require $M_1 \geq 25 M_\odot$. BH formation from core collapse generally requires a progenitor with ZAMS mass $M_{\rm ZAMS} \geq 20 M_\odot$ (e.g., \citealt{Belczynskietal2016,MandelFarmer2022}). The adopted threshold of $25 M_\odot$ lies conservatively above this minimum, ensuring that the progenitor possesses a sufficiently large core binding energy for BH formation across a range of remnant-mass prescriptions. 

\begin{figure}[]
\centering
\includegraphics[width=\columnwidth]{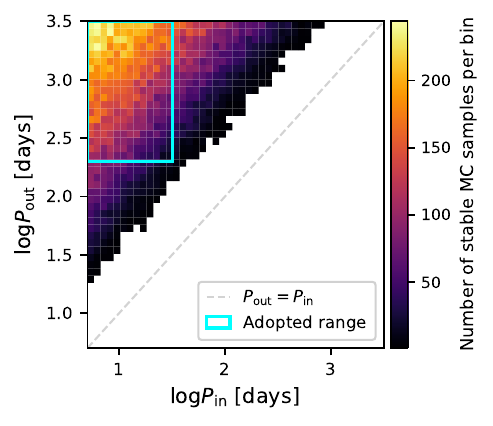}
\caption{Probability density of dynamically stable hierarchical triples in the ($\log P_{\mathrm{in}}, \log P_{\mathrm{out}}$) plane at ZAMS. The orbital period $P$ is reported in days. The light blue rectangle marking the adopted ($\log P_{\mathrm{in}}, \log P_{\mathrm{out}}$) range: $\log P_{\mathrm{in}} \in [0.7,1.5]$ and $\log P_{\mathrm{out}} \in [2.3,3.5]$. The color scale shows the number of dynamically stable Monte Carlo samples per bin, out of 500,000 systems drawn from the \cite{MoeDiStefano2017} joint distributions; stability is evaluated as a binary criterion via \cite{AarsethMardling2001}, so the density reflects how stable triples concentrate in the ($\log P_{\rm in}, \log P_{\rm out}$) plane.}
\label{fig:density_map}                     
\end{figure}

\begin{figure*}[]
\centering
\includegraphics[width=\textwidth]{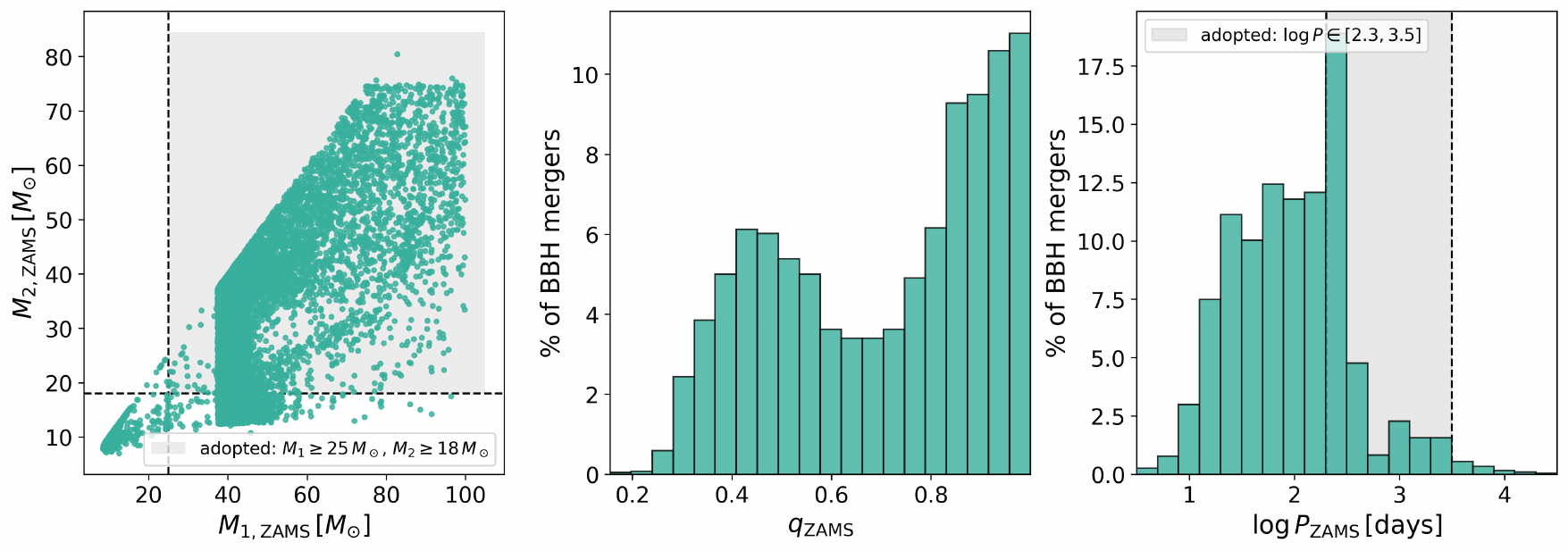}
\caption{
Initial ZAMS property distributions for binary systems that evolve into merging BBHs. The data was extracted from the \textsc{compas} population synthesis simulations by \cite{Grichener2023}. These distributions serve as a guide for constraining the initial parameters of our model. 
Left panel: ZAMS masses of the BH progenitors. Middle panel: initial mass ratio distribution, $q_{\mathrm{ZAMS}} \equiv M_{2,\mathrm{ZAMS}}/M_{1, \mathrm{ZAMS}}$ that leads to BBH mergers. The turquoise bins illustrate the percentage contribution of each mass-ratio bin relative to the total merging BBH population (the columns in the histogram add up to 100 \%). The right panel: similar to the middle panels for the $\log P_{\mathrm{ZAMS}}$.
}
\label{fig:Histograms_compas}               
\end{figure*}

\textbf{Orbital period.} The orbital period is restricted to $\log(P/\text{day}) \in [2.3, 3.5]$; 
The upper bound lies well within the interaction regime identified by \cite{MoeDiStefano2017}, who find that massive binaries with $\log(P/\text{day}) \lesssim 3.7$ undergo Roche-lobe overflow during post-main-sequence evolution. This ensures that the binary undergoes mass transfer and/or CEE as required by the evolutionary scenario described in section \ref{sec:assumptions}. 
The lower bound of $\log(P/\text{day}) = 2.3$ is adopted to coincide with the outer-orbit period range used in the triple calculation of section \ref{subsec:TripleResults}, which is itself set by the hierarchical-stability requirement of \cite{AarsethMardling2001}; see Figure \ref{fig:density_map}. Matching the binary period range to the triple outer-orbit range makes the binary and triple draw probabilities directly comparable; in the conditional-probability decomposition adopted in this paper, the binary calculation can then be read as the outer-orbit sub-case of the triple calculation. We note that this lower bound is significantly more conservative than the threshold for main-sequence interaction in massive binaries (e.g., \citealt{Marchantetal2016, Wellsteinetal2001, Klenckietal2026}), which becomes important only at $\log(P/\text{day}) \lesssim 1$; the systems excluded by our cut at $1 \lesssim \log(P/\text{day}) < 2.3$ are therefore physically viable for post-MS interaction but are removed for the comparability reason stated above.
As a cross-check, we examined the ZAMS properties of all systems that produce BBH mergers in the \textsc{compas} (Compact Object Mergers: Population Astrophysics and Statistics; \citealt{Stevensonetal2017, VignaGomezetal2018, TeamCOMPAS2022} version 02.31.06 \citealt{Grichener2023}\footnote{Publicly available at Zenodo: \href{https://zenodo.org/records/11237180}{doi:10.5281}.}), population synthesis dataset shown in Figure \ref{fig:Histograms_compas}. The orbital period distribution of merging systems is concentrated at $\log(P/\text{day}) \sim 1.5-3$, with the dominant peak near $\log(P/\text{day}) \sim 2-2.5$ and very few systems below $\log(P/\text{day}) = 1$. The adopted range [2.3,3.5] thus lies on the high-period flank of the merging-system distribution and is conservative relative to its peak. We note that \textsc{compas} employs a BH formation prescription based on pre-collapse core mass and remnant-mass mappings, in which a sufficiently massive star can form a BH without the rapid core rotation required by the JJEM. The \textsc{compas} - derived distributions are therefore not directly applicable to the present scenario, and are used here only to verify that the physically motivated ZAMS cuts are not implausibly narrow or wide.
%FFFFFFFFFFFFFFFFFFFFFFFFFFFFFFFFFFFFFF
%  trim={<left> <lower> <right> <upper>}
%FFFFFFFFFFFFFFFFFFFFFFFFFFFFFFFFFFFFFF
%FFFFFFFFFFFFFFFFFFFFFFFFFFFFFFFFFFFFFF
%  trim={<left> <lower> <right> <upper>}
%FFFFFFFFFFFFFFFFFFFFFFFFFFFFFFFFFFFFFF

\textbf{Secondary mass.} We require $M_2 \geq 18 M_\odot$ at ZAMS. BH formation in standard remnant-mass prescriptions requires a ZAMS-equivalent mass of approximately $20 M_\odot$ or more (e.g., \citealt{Belczynskietal2016,MandelFarmer2022}). A ZAMS threshold below this value is adopted because the evolutionary scenario \ref{sec:assumptions} includes mass transfer from the primary RSG to the secondary prior to the first core collapse. As noted by \cite{Soker2022misalignment}, this accretion can substantially increase the secondary mass above its birth value, potentially even exceeding $M_{1,\rm ZAMS}$. 
The relevance of this channel rests on the well-established observation that massive stars rarely evolve in isolation: \cite{Sanaetal2012} find that more than $70 \%$ of O-type stars are members of close interacting binaries that exchange mass with a companion during their evolution. 
Detailed population studies further show that a substantial fraction of these massive binaries produce accretors whose mass is significantly increased through stable Roche-lobe overflow from the primary, with the gain scaling with the donor's envelope mass and the assumed accretion efficiency (\citealt{DeMinketal2014}). On this basis, accretion of at least $2 M_\odot$ onto a $18 M_\odot$ secondary - sufficient to bring its pre-collapse mass above the BH-formation threshold - is a conservative expectation. 
As illustrated in the left and middle panels of Figure \ref{fig:Histograms_compas}, approximately 26\% of the BBH-merging systems within the \textsc{compas} population feature $M_{1,\mathrm{ZAMS}} < 18 M_\odot$, while 40\% have secondary masses below this same threshold ($M_{2,\mathrm{ZAMS}} < 18 M_\odot$). This demonstrates that our adopted thresholds of $25 M_\odot$ for the primary and $18 M_\odot$ for the secondary are conservative: a substantial fraction of \textsc{compas} mergers begin below them, so relaxing the thresholds would only increase $P_{\mathrm{draw}}$ (and lower the required $\beta$). 
The adopted threshold thus serves as a ZAMS proxy for the physical requirement on the pre-collapse mass.

\textbf{Eccentricity.} The eccentricity is drawn from the power-law distribution described in section \ref{subsec:outline} (\citealt{MoeDiStefano2017}), without additional range restriction beyond the Roche-lobe-filling criterion in the periastron. Because this is the only constraint imposed on $e$, the eccentricity is sampled freely over the physically allowed interval $[0, e_{\max}(P)]$ for each configuration $(M_1, \log P, \ q)$, and the integration of the joint probability density function over this interval contributes only a normalization factor. Eccentricity, therefore, does not impose a conditional restriction on $P_{\mathrm{draw}}$ beyond what is absorbed in this normalization.

Integrating the empirically calibrated joint distribution (equation \ref{eq:prob_expression}) over this region yields a fiducial binary draw probability $P_{\mathrm{draw}} \simeq 3 \times 10^{-2}$. The constituent conditional probabilities that contribute to this aggregate value are detailed below. For consistency, all subsequent mass and temporal quantities are reported in solar masses ($M_\odot$) and days, respectively.

\begin{subequations}
\label{eq:prob_binary}
\begin{align}
& P\!\left(M_1 \in [25,150] \mid M_1 \ge 8\right) && \simeq 0.199, \label{eq:prob_binary_m1}\\[7pt]
& P\!\left(n > 0 \mid M_1 \in [25,150]\right) && \simeq 0.915, \label{eq:prob_binary_n}\\[7pt]
& \begin{aligned}
    &P\bigl(\log P \in [2.3,3.5] \mid n>0, \\[1pt]
    &\quad M_1 \in [25,150]\bigr)
  \end{aligned}
  && \simeq 0.375, \label{eq:prob_binary_logP}\\[7pt]
& \begin{aligned}
    &P\bigl(M_2 \in [18,M_1] \mid \log P \in [2.3,3.5], \\[1pt]
    &\quad n>0,\, M_1 \in [25,150]\bigr)
  \end{aligned}
  && \simeq 0.441, \label{eq:prob_binary_m2}
\end{align}
\end{subequations}
where n is the number of companions of the primary with $q \geq 0.1$, counted over the full empirical period range $0.2 \leq \log P \leq 8.0$.

When inserted into the CCSN-normalized rate relation (equation \ref{eq:beta}), this corresponds to a success fraction $\beta_{\mathrm{binary}} \simeq 0.023$.
This value is physically plausible: it is consistent with the expectation that only a small fraction of initially favorable systems survive the combined uncertainties of CE ejection efficiency, natal kick disruption, and the requirement of merger within a Hubble time. 

A further effect that reduces the number of merging-BBH systems is the requirement of a tertiary star (Section \ref{sec:assumptions}), which we examine next.

% =========================
\subsection{Triple calculation}
\label{subsec:TripleResults}
% =========================

We extend the calculation to hierarchical triples by introducing an additional inner companion to the primary - the tertiary star of mass $M_{3}$ (Figures \ref{fig:evolution_channel} and \ref{fig:am_geometry}).

\textbf{Outer orbit.} The outer orbit is retained within the same intervals $\log(P_{\text{in}}/\text{day}) \in [2.3, 3.5]$ as in the binary case, and the outer companion mass is required to satisfy $M_{2} \in [18,150] M_\odot$, (with the implicit constraint $M_{2} < M_{1}$), for the same physical reasons discussed in section \ref{subsec:BinaryResults}. The outer-orbit parameters therefore contribute the same conditional probabilities as in the binary calculation. 

\begin{figure}
\centering
\includegraphics[width=\columnwidth]{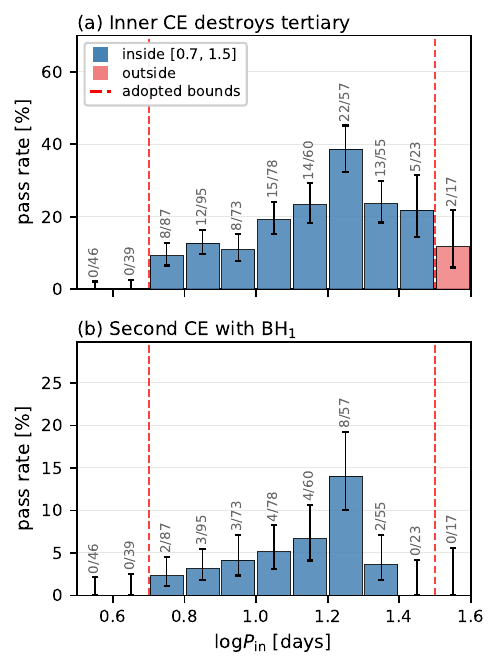}
\caption{Simulated justification of the adopted inner-orbit period range $\log(P_{\text{in}}/\text{day}) \in [0.7,1.5]$ from two MSE runs. The first run samples $\log(P_{\text{in}}/\text{day})$ within $[0.5,1.4]$ and the second run samples $\log(P_{\text{in}}/\text{day})$ within $[0.7,1.6]$ (days), overall $N=630$. Panel (a): fraction of systems per 0.1-dex bin that initiate the inner CEE, in which the primary engulfs and tidally destroys the tertiary. Panel (b): fraction that subsequently reaches the second CEE, in which the first-born black hole enters a second common-envelope phase with the secondary. Error bars give $68\%$ Wilson binomial confidence intervals. The counts above each bar denote (number of passers) / (number sampled). Vertical dashed lines mark the bounds adopted for the analytic calculation; bars in lightcoral lie outside this range. No system below $\log(P_{\text{in}}/\text{day})=0.7$ initiates the inner common envelope at all, motivating the lower bound; the second-CEE fraction drops to zero. above $\log(P_{\text{in}}/\text{day}) \approx 1.4$, motivating the upper bound.}
\label{fig:fig_logPin_bounds}                
\end{figure}

\textbf{Inner orbit.} The inner orbital period is restricted to $\log(P_{\text{in}}/\text{day}) \in [0.7,1.5]$. This choice is motivated by several considerations. This interval lies within the empirically established interaction regime for massive binaries (e.g., \citealt{MoeDiStefano2017}): the tertiary must be engulfed once the primary expands after the MS, which requires a sufficiently tight inner orbit. Both bounds are motivated by the time evolution runs with the Multiple Stellar Evolution (MSE) code (\citealt{Hamersetal2021}) and have distinct physical origins (Figure \ref{fig:fig_logPin_bounds}). The lower bound is set by the onset of the inner CEE: at $\log(P_{\text{in}}/\text{day}) < 0.7$ the inner orbit is so tight that the primary fills its Roche lobe, or otherwise interacts with the tertiary, while both stars are still on the MS - that is, before the primary expands after the MS and drives the inner CEE of the channel. In the pilot runs, no system below $\log(P_{\text{in}}/\text{day}) = 0.7$ initiates the inner CEE at all (Figure \ref{fig:fig_logPin_bounds}a). The upper bound is set, by contrast, by the failure of the second CEE. The ZAMS stability criterion alone (equation \ref{eq:stability}) would permit inner periods up to $\log(P_{\text{in}}/\text{day}) \simeq 1.85$, but the pilot runs show the channel ceasing to complete well below this value: the fraction of systems that proceed to the second CEE between the first-born BH and the secondary declines with increasing inner period and drops to zero above $\log(P_{\text{in}}/\text{day}) \simeq 1.4-1.5$ (Figure \ref{fig:fig_logPin_bounds}b). Such wide-inner-orbit triples can still undergo the first CEE but do not complete the remainder of the sequence, in part because a wider inner orbit leaves the system more susceptible to dynamical instability as the stars evolve. The value $\log(P_{\text{in}}/\text{day}) = 1.5$ is therefore adopted as the period above which the channel is no longer realized in the simulations, rather than as the formal ZAMS-stability limit. Both bounds thus reflect the evolution of the triple and not merely its zero-age configuration: the ZAMS stability criterion is necessary but not sufficient, and the time-evolution runs localize the inner-period range over which the full channel proceeds.
%FFFFFFFFFFFFFFFFFFFFFFFFFFFFFFFFFFFFFF
%  trim={<left> <lower> <right> <upper>}
%FFFFFFFFFFFFFFFFFFFFFFFFFFFFFFFFFFFFFF

\textbf{Inner companion mass.} The inner companion (tertiary) mass is required to satisfy $M_3 \in [1.25,4.5] M_\odot$. The upper bound reflects the requirement $M_3 < 5 M_\odot$ from the evolutionary scenario (Section \ref{sec:assumptions}): the tertiary must be a low-mass star that can be tidally disrupted by the primary's core during the first CEE, rather than a massive star that would itself significantly affect the system's evolution prior to being engulfed. The lower bound is not an independent mass cut but a consequence of the sampling: it follows from the mass-ratio floor $q_{\min}=0.05$ the extrapolation limit of the \cite{MoeDiStefano2017} distribution adopted in section \ref{subsec:zams} - together with the minimum primary mass $M_1 = 25 M_\odot$, so that $M_{\mathrm{in}} \geq q_{\min} \times M_{1,\min} = 1.25 M_\odot$. The corresponding inner mass ratios $q_{\mathrm{in}} = M_{\mathrm{in}}/M_{1}$ thus span approximately $0.05-0.3$ for primaries $M_1 \geq 25 M_\odot$, lying at the low end of - and partially below - the calibrated domain of \cite{MoeDiStefano2017} ($q \geq0.1$). Tertiaries near this lower bound carry only a small orbital angular momentum relative to the primary; the spin-up of the primary's core required by the JJEM (Section \ref{sec:assumptions}) is therefore expected to be most effective for tertiaries toward the upper part of the adopted mass range.

\textbf{Dynamical stability.} The \cite{AarsethMardling2001} stability criterion (equation \ref{eq:stability}) is imposed as a hard cut in the ZAMS parameter space. The fraction of Monte Carlo draws that satisfy the stability criterion, among systems that already meet all other constraints, is $P_{\rm stable} \simeq 0.58$.

Within this fiducial triple region, the conditional probabilities contributing to the draw probability are: (the following mass and temporal quantities are in units of solar masses ($M_\odot$) and days, respectively).

\begin{subequations}
\label{eq:prob_triple}
\begin{alignat}{2}
& P\!\left(M_1 \in [25,150] \mid M_1 \ge 8\right) && \simeq 0.199, \label{eq:prob_triple_m1}\\[7pt]
& P\!\left(n_{\mathrm{in}} > 0 \mid M_1 \in [25,150]\right) && \simeq 0.961, \label{eq:prob_triple_n_in}\\[7pt]
& P\!\left(n_{\mathrm{out}} > 0 \mid M_1 \in [25,150]\right) && \simeq 0.915, \label{eq:prob_triple_n_out}\\[7pt]
& \begin{aligned}
    &P\bigl(\log P_{\mathrm{in}} \in [0.7,1.5] \mid n_{\mathrm{in}}>0, \\[1pt]
    &\quad M_1 \in [25,150]\bigr)
  \end{aligned}
  && \simeq 0.283, \label{eq:prob_triple_logP_in}\\[7pt]
& \begin{aligned}
    &P\bigl(\log P_{\mathrm{out}} \in [2.3,3.5] \mid n_{\mathrm{out}}>0, \\[1pt]
    &\quad M_1 \in [25,150]\bigr)
  \end{aligned}
  && \simeq 0.375, \label{eq:prob_triple_logP_out}\\[7pt]  
& \begin{aligned}
    &P\bigl(M_3 \in [1.25,4.5] \mid \log P_{\mathrm{in}} \in [0.7,1.5], \\[1pt]
    &\quad n_{\mathrm{in}}>0,\, M_1 \in [25,150]\bigr)
  \end{aligned}
  && \simeq 0.221, \label{eq:prob_triple_m3}\\[7pt]
& \begin{aligned}
    &P\bigl(M_2 \in [18,M_1] \mid \log P_{\mathrm{out}} \in [2.3,3.5], \\[1pt]
    &\quad n_{\mathrm{out}}>0,\, M_1 \in [25,150]\bigr)
  \end{aligned}
  && \simeq 0.441, \label{eq:prob_triple_m2}  
\end{alignat}
\end{subequations}
where $n_{\mathrm{in}}$ and $n_{\mathrm{out}}$ are the number of companions of the primary with $q \geq 0.05$ and $q \geq 0.01$, respectively, counted over the full empirical period range $0.2 \le \log P \le 8.0$. 

Expressed as a fraction of the massive-primary population, the calculation implies that roughly $0.9 \%$ of primaries with $M_1 \geq 25 M_\odot$ host companions at both the inner and outer separations of interest, of which roughly half form dynamically stable configurations - a modest but non-negligible fraction.

Inserting $P_{\rm draw}^{\rm triple}$ into the CCSN-normalized rate relation (equation \ref{eq:beta}) gives the success fraction for the triple channel $\beta_{\mathrm{triple}} \lesssim 0.67$. 
As in the binary case, this value is best read as an upper bound: it is the per-system success fraction the channel would require if it produced the entire observed BBH merger rate, whereas in practice it contributes alongside other formation channels. 
This bound should be interpreted cautiously, because $\beta_{\mathrm{triple}}$ is high - about thirty times the corresponding binary value ($\beta_{\mathrm{binary}} \simeq 0.023$), the difference reflecting the multiplicity penalty that makes suitable triples rarer. Producing the entire observed rate through this channel would require roughly four in five of the suitable triples to complete the full sequence of section \ref{sec:assumptions} - the inner CEE and tidal disruption of the tertiary, two core collapses each forming a BH, a second CEE that leaves a bound and sufficiently tight orbit, and coalescence within a Hubble time - despite the large uncertainties that $\beta$ absorbs, including CE ejection efficiency, natal kicks, mass loss, and the JJEM core-rotation requirement that the simulations do not test (section \ref{subsec:mse}). A success fraction this large is demanding for so long and uncertain an evolutionary chain. We therefore do not read $\beta_{\mathrm{triple}} \simeq 0.67$ as evidence that the triple channel produces the whole BBH population; the more natural reading is that it is a viable contributor supplying some fraction of the observed rate, in which case the required per-system success fraction falls proportionally to a more comfortable value. The bound nonetheless lies below unity, so the channel is not excluded on rate grounds and is promising as a contributor to the merging-BBH population -  but the present work does not establish that it dominates that population.

% =========================
\subsection{Expanding the boundaries of the constraints}
\label{subsec:exp_constraints}
% =========================

The fiducial primary-mass cut, $M_1 \geq 25\,M_\odot$, was placed conservatively
above the $\simeq 20\,M_\odot$ minimum generally required for BH formation (section \ref{subsec:BinaryResults}). To gauge how sensitive the inferred success
fraction is to this choice, we repeat the triple calculation with the primary mass
relaxed to the BH-formation floor, $M_1 \in [20,150]\,M_\odot$. 
Admitting lower-mass primaries enlarges the qualifying population and thus raises the draw probability; since $\beta \propto P_{\rm draw}^{-1}$ (equation \ref{eq:beta}), this can only reduce the required success fraction. The conditional probabilities for the
relaxed cut are: (the following mass and temporal quantities are in units of solar masses ($M_\odot$) and days, respectively).

\begin{subequations}
\label{eq:prob_triple_20}
\begin{alignat}{2}
& P\!\left(M_1 \in [20,150] \mid M_1 \ge 8\right) && \simeq 0.276, \label{eq:prob_triple_m1_20}\\[7pt]
& P\!\left(n_{\mathrm{in}} > 0 \mid M_1 \in [20,150]\right) && \simeq 0.953, \label{eq:prob_triple_n_in_20}\\[7pt]
& P\!\left(n_{\mathrm{out}} > 0 \mid M_1 \in [20,150]\right) && \simeq 0.9, \label{eq:prob_triple_n_out_20}\\[7pt]
& \begin{aligned}
    &P\bigl(\log P_{\mathrm{in}} \in [0.7,1.5] \mid n_{\mathrm{in}}>0, \\[1pt]
    &\quad M_1 \in [20,150]\bigr)
  \end{aligned}
  && \simeq 0.267, \label{eq:prob_triple_logP_in_20}\\[7pt]
& \begin{aligned}
    &P\bigl(\log P_{\mathrm{out}} \in [2.3,3.5] \mid n_{\mathrm{out}}>0, \\[1pt]
    &\quad M_1 \in [20,150]\bigr)
  \end{aligned}
  && \simeq 0.367, \label{eq:prob_triple_logP_out_20}\\[7pt]  
& \begin{aligned}
    &P\bigl(M_3 \in [1.25,4.5] \mid \log P_{\mathrm{in}} \in [0.7,1.5], \\[1pt]
    &\quad n_{\mathrm{in}}>0,\, M_1 \in [20,150]\bigr)
  \end{aligned}
  && \simeq 0.307, \label{eq:prob_triple_m3_20}\\[7pt]
& \begin{aligned}
    &P\bigl(M_2 \in [18,M_1] \mid \log P_{\mathrm{out}} \in [2.3,3.5], \\[1pt]
    &\quad n_{\mathrm{out}}>0,\, M_1 \in [20,150]\bigr)
  \end{aligned}
  && \simeq 0.355. \label{eq:prob_triple_m2_20}  
\end{alignat}
\end{subequations}

These factors give an outer draw probability $P_{\rm out} \simeq 3.3\times10^{-2}$ and an inner conditional probability $P_{\rm in} \simeq 7.8\times10^{-2}$, while the fraction of stability-satisfying configurations is $P_{\rm stable} \simeq 0.57$. Expressed as a fraction of the massive-primary population, roughly $0.9\%$ of primaries with $M_1 \geq 20\,M_\odot$ host companions at both the inner and outer separations of interest, of which slightly more than half remain dynamically stable. The total draw probability is therefore $P_{\rm draw}^{\rm triple} \simeq 1.5\times10^{-3}$ (equation \ref{eq:triplefactor}), which, inserted into the CCSN-normalized relation (equation \ref{eq:beta}), gives $\beta_{\rm triple} \simeq 0.47$.

Relaxing the primary-mass cut to the BH-formation floor thus lowers the required success fraction from the fiducial $\beta_{\rm triple} \simeq 0.67$ (section \ref{subsec:TripleResults}) to $\simeq 0.47$. The channel's viability does not hinge on the precise primary-mass threshold: a wider primary range only moves the inferred $\beta$ further below unity.

% ==================================
\section{Summary} 
\label{sec:Summary}
% ==================================

GW observations show that the distribution of the effective inspiral spin parameter $\chi_{\mathrm{eff}}$  (equation \ref{eq:EffectiveSpin}) of merging BBHs is asymmetric about zero: it peaks slightly positive but retains a significant negative tail (Figure \ref{fig:chieff_distribution}), with an inferred negative fraction of $0.3-0.46$ (\citealt{2026arXiv260527226T}). Neither isolated binary evolution, which predicts $\chi_{\mathrm{eff}}$ narrowly peaked near zero, nor gas-free dynamical assembly, which predicts a distribution symmetric about zero, reproduces this asymmetry on its own (Section \ref{sec:intro}). 
  
In this paper, we examined whether a hierarchical triple-star channel, which we formulated within the framework of the JJEM, can contribute a non-negligible fraction of the misaligned population. In this channel (Figures \ref{fig:evolution_channel} and \ref{fig:am_geometry}), the primary engulfs a low-mass tertiary in a first CEE, whose inspiral spins up the primary's core and enables BH formation according to the JJEM. The secondary subsequently engulfs the first-born BH in a second CEE and collapses to the second-born BH (Section \ref{sec:assumptions}). The spin–orbit misalignment arises from the inclination between the inner and outer orbital planes, rather than from a natal kick or from purely dynamical forcing. Namely, the misalignment is inherited from the formation of the triple system. 

We set constraints on the parameters of the triple system (equation \ref{eq:prob_triple}): the masses of the three stars, the inner and outer orbital periods, and the eccentricity. We demanded that during the main-sequence phase of the primary star, the triple system be dynamically stable. We semi-analytically estimated the expected fraction of merging BBHs under our proposed scenario, and determined its ratio to the CCSN explosion rate (equation \ref{eq:rate}). Comparing this ratio to the observed one, we found the fraction $\beta$ of systems obeying our initial criteria that should end as observed merging BBH. 
For our fiducial triple-star parameters, we found that a fraction of $\beta_{\mathrm{triple}} \simeq 0.67$ of our initial systems should end as merging BBH. As this value lies below unity, the evolutionary channel we propose might contribute significantly to merging BBHs. Namely, the rate does not exclude this channel from forming most or all of the merging BBHs. Relaxing the constraints somewhat, like allowing a lower minimum mass for the primary star (equation \ref{eq:prob_triple_20}), eases the formation of systems and reduces the value of $\beta_{\mathrm{triple}}$.      

Our proposed channel makes an interesting prediction that distinguishes it from dynamical interaction channels. In our proposed channel, the first BH to form is mostly misaligned with the orbital angular momentum. The second BH is mostly aligned with the orbital angular momentum. We therefore predict that there is a subpopulation in the merging BBH population that has one BH with an almost random spin with respect to the orbital angular momentum, and one that is almost aligned with the orbital angular momentum. In the dynamical interaction channels, both spins are mostly random. The first BH to form is likely to be the more massive one, but in a large fraction of systems the second BH to form might be the more massive one. Therefore, the mass of the BH is not the best parameter to distinguish between the first BH to form and the second one. The spin direction is the better one in the channel we proposed here.    

The main advantage of our proposed channel is that it produces both a positive average $\chi_{\mathrm{eff}}$ and a significant negative tail (Section \ref{sec:assumptions}). Therefore, it can qualitatively reproduce the main features of the observed distribution and is a promising non-negligible contributor to the misaligned population of merging BBHs. Whether it dominates this population or contributes less than half is a subject for future study. As well, there is a need to derive the quantitative distribution of $\chi_{\mathrm{eff}}$. 

% ======================================
\section*{Acknowledgements}
% ======================================

We thank an anonymous referee for helpful comments. 
NS thanks the Charles Wolfson Academic Chair at the Technion for the support.

% =================================
% =================================

% =================================
% =================================
% =================================

%%% Below is for using the bib file

% =================================
% =================================
% =================================

%\bibliographystyle{mnras}
 \bibliography{reference}{}
  \bibliographystyle{aasjournal}

% --- Appendix Section ---
\appendix
\section{Conversion of the BBH merger rate to z=0}
\label{app:rate_conversion}

\cite{2026arXiv260527226T} reports the density of the BBH merger-rate at $z=0.2$ rather than at $z=0$. Because our normalization compares this quantity with the local ($z \simeq 0$) CCSNe rate (\citealt{Maetal2025}), the two must be evaluated in the same redshift; therefore, we convert the GWTC-5.0 value to $z=0$.

GWTC-5.0 models the redshift distribution of mergers as
\begin{equation}
\pi(z\mid\kappa_z) \;\propto\; \frac{1}{1+z}\,\frac{dV_c}{dz}\,(1+z)^{\kappa_z},
\label{eq:piz}
\end{equation}
where $dV_c/dz$ is the comoving volume element and the factor $1/(1+z)$
accounts for the time dilation between the source and detector frames. The
astrophysical content of this model is carried out by the comoving merger-rate density, which scales as $\mathcal{R}(z) \propto (1+z)^{\kappa_z}$; this is the quantity tabulated by GWTC-5.0 and the one relevant to our normalization. The remaining factors in equation \ref{eq:piz} convert this rate density into the observed distribution of detections and do not enter the rescaling of the rate density between redshifts. 
Because $\mathcal{R}(z)\propto(1+z)^{\kappa_z}$, the ratio of the rate at two redshifts is independent of its overall normalization,
\begin{equation}
\frac{\mathcal{R}(z=0)}{\mathcal{R}(z=0.2)}
  = \left(\frac{1+0}{1+0.2}\right)^{\kappa_z}
  = (1.2)^{-\kappa_z},
\end{equation}
so that
\begin{equation}
\mathcal{R}(z=0) = \mathcal{R}(z=0.2)\,(1.2)^{-\kappa_z},
\label{eq:conv}
\end{equation}

GWTC-5.0 infers $\kappa_z = 2.5\pm0.7$ for the BBH population, consistent
with the redshift evolution of the star-formation rate (\citealt{MadauDickinson2014}). %consistent with the low-redshift slope of the cosmic star-formation rate
With $\kappa_z=2.5$ the conversion factor is $(1.2)^{-2.5} \simeq 0.633$ and equation \ref{eq:conv} maps the reported interval $27.5-49.4 \ \mathrm{Gpc^{-3}\,yr^{-1}}$ to 
\begin{equation}
\mathcal{R}_{\rm BBH}(z=0) \simeq 17.4-31.3 \ \mathrm{Gpc^{-3}\,yr^{-1}}.
\end{equation}
The uncertainty on $\kappa_z$ propagates only weakly: over
$\kappa_z\in[1.8,3.2]$ the factor ranges from $(1.2)^{-1.8} \simeq 0.72$ to
$(1.2)^{-3.2} \simeq 0.55$, a $\sim 15\%$ spread that is subdominant to the
width of the rate interval itself.
As an independent check, the converted value is consistent with the $z=0$ BBH rate reported directly by the earlier GWTC-4.0 analysis, $\mathcal{R}_{\mathrm{BBH}} = 14-26 \ \mathrm{Gpc^{-3}\yr^{-1}}$ (\citealt{LVK2025GWTC4pop}).

\end{document}